# Coupled breathing modes in one-dimensional Skyrmion lattices

Junhoe Kim, Jaehak Yang, Young-Jun Cho, Bosung Kim, and Sang-Koog Kim [a]

*National Creative Research Initiative Center for Spin Dynamics and Spin-Wave Devices, Nanospinics Laboratory, Research Institute of Advanced Materials, Department of Materials Science and Engineering, Seoul National University, Seoul 151-744, Republic of Korea*

We explored strong coupling of dynamic breathing modes in one-dimensional (1D) skyrmion lattices periodically arranged in thin-film nanostrips. The coupled breathing modes exhibit characteristic concave-down dispersions that represent the in-phase high-energy mode at zero wavenumber ($k$=0) and the anti-phase low-energy mode at the Brillouin zone boundary ($k=k_{BZ}$). The band width of the allowed modes increases with decreasing inter-distance between nearest-neighboring skyrmions. Furthermore, the collective breathing modes propagate very well through the thin-film nanostrips, as fast as 200 ~ 700 m/s, which propagation is controllable by the strength of magnetic fields applied perpendicularly to the film plane. The breathing modes in 1D skyrmion lattices potentially formed in such nanostrips possibly can be used as information carriers in information processing devices.

a) Author to whom all correspondence should be addressed; electronic mail: sangkoog@snu.ac.kr

## I. INTRODUCTION

Magnetic skyrmions[1,2] are topologically stable spin textures as found in bulk crystals[3,4] and ultrathin-film layered structures[5-7] owing to a broken inversion symmetry and a strong spin-orbit coupling. Dzyaloshinskii-Moriya interaction (DMI)[8] in these systems plays a key role in the formation of magnetic skyrmions[3-7]. The exotic topological spin texture of skyrmions and their manipulation, as well as their dynamic modes in the high-frequency sub-GHz-to-several-tens-of-GHz range are of increasing interest from both fundamental and technological perspectives[6,9-23]. For example, reliable control of skyrmion motions in narrow-width thin-film nanostrips by spin-polarized currents or magnetic fields allows for implementation of skyrmions in potential information-storage and -processing devices[6,9-14]. Moreover, single skyrmions in geometrically restricted magnetic dots exhibit unique GHz-range dynamic motions such as lower-energy gyration modes and higher-energy breathing modes[15-17]. Therefore, microwave generators and detectors have been proposed based on their inherent dynamic modes[18-21]. Very recently, the collective gyration modes in coupled individual skyrmions have been explored as information carriers[22,23], as analogous to propagating spin waves in magnonic crystals[24-28] and to propagating gyration modes in physically connected or separated vortex-state disks[29-31].

In this work, we explore the coupled breathing modes between the nearest-neighboring skyrmions in thin-film nanostrips showing characteristic concave-down dispersions. The collective motions of the individual skyrmions' breathing modes propagate well through the continuous nanochannels owing to the strong coupling characteristics of neighboring skyrmions. Their propagation speed, in fact, is much higher than those of skyrmion arrays' coupled gyration modes[23], and also can be controlled by applying perpendicular magnetic fields.

## II. MODELING AND SIMULATIONS

In the present study, we used the Mumax3 code[32] that incorporates the Landau-Lifshitz-Gilbert (LLG) equation[33] to numerically calculate the dynamic motions of individual skyrmions periodically arranged in magnetic nanostrips. Co thin film interfaced with Pt was modeled as follows[9]: saturation magnetization $M_s$ = 580 kA/m, exchange stiffness $A_{ex}$ = 15 pJ/m, perpendicular anisotropy constant $K_u$ = 0.8 MJ/m$^3$, DMI constant $D$ = 3.0 mJ/m$^2$. The unit-cell size used in the micromagnetic simulations was $1.0 \times 1.0 \times 1.0$ nm$^3$. An initial magnetization state was assumed to be that of the Néel-type skyrmion, which was then relaxed for 100 ns using the damping constant α = 0.3, until its ground state could be obtained. Once we had obtained the ground state of single skyrmions or skyrmion lattices, we used the damping constant $α$ = 0.0001 to excite the dynamic modes for better spectral resolution in our analyses.

## III. RESULTS AND DISCUSSION

First, in order to excite the breathing mode of a single skrymion in a square dot as depicted in Fig. 1(a), we applied the sinc-function field $H_z = H_0 \sin[\omega_H(t-t_0)]/[\omega_H(t-t_0)]$ for 100 ns with $H_0 = 10 \, \text{Oe}$, $\omega_H = 2\pi \times 50 \, \text{GHz}$, and $t_0$= 1 ns. The temporal oscillation of <$m_z$>, $m_z$ averaged over the entire region and its fast Fourier transformation (FFT) spectrum are plotted in Figs. 1(b) and 1(c), respectively. A single peak appeared at 22.32 GHz, as correspondent with the breathing-mode frequency. Figure 1(d) shows the temporal oscillation $<\Delta m_z(t)> (= <m_z(t)> - <m_z(t=0)>)$ (left column) along with snapshot images of the local $\Delta m_z$ (right column), as obtained by the inverse FFT of the FFT power of <$m_z$> and the local $m_z$ in the frequency regions of $\Delta f$ =22.30-22.34 GHz, respectively. The $m_z$ oscillation in the core region with its maintained radial symmetry corresponds to a characteristic breathing-mode behavior representative of periodic expansion and contraction of the core[15-17].

Next, we examined two skyrmions placed in a rectangular dot as indicated in Fig. 2(a). To excite all of the possible breathing modes of the two skyrmions, we applied, in the $-z$ direction, a 10 ns-width pulse field of 500 Oe locally to only one of the two skyrmions. After the local field was turned off, we monitored, under free relaxation for 100 ns, the $<m_z>$ in two different regions surrounding each core (*i.e.* $x$=0-32 nm for the left skyrmion and $x$ = 33-64 nm for the right skyrmion), as shown in Fig. 3(b). The FFTs of the $<m_z>$ oscillations in the two regions are plotted in Fig. 3(c). Two distinct peaks denoted $\omega_l$ and $\omega_h$ are shown at $2\pi \times 20.55$ and $2\pi \times 24.66$ GHz, respectively, in the given frequency range[34]. The frequency splitting for the two skyrmions was the result of the symmetry breaking of the potential energy profile of the isolated skyrmion due to its coupling, similarly to the coupled vortex disks described in refs. 35 and 36. From the inverse FFTs of the local $m_z$, we obtained, in the perspective view, snapshot images of both the lower $\omega_l$ and higher $\omega_h$ modes. For the $\omega_l$ mode, the left-side and right-side skyrmion cores oscillate as a breathing mode; that is, each core expands and contracts periodically, but both cores oscillate in anti-phase with each other. On the other hand, for the $\omega_h$ mode, both cores oscillate in-phase, as shown in Fig. 2(d). The anti-phase breathing motion of the two skyrmions shows a lower energy than the in-phase breathing mode, because the breathing modes represent the expansion and contraction of skyrmion cores. Therefore, the expansion of both cores in the same $z$ (or $-z$) direction incurs higher energy cost of dynamic periodic motion. This effect is opposite to those of the coupled gyration modes of skyrmions or magnetic vortices[23,29]. In-phase gyration motion between neighboring skyrmions or magnetic vortices shows a lower energy than anti-phase gyration motion. This can be explained by the dynamic dipolar interaction between the gyrations of neighboring disks in vortex-state disks[29], while the exchange, DMI, and magnetic anisotropy, as well as the dipolar energy together contribute to the coupled gyration modes of neighboring skyrmions[23].

Additionally, we studied with a coupling of five skyrmions in a nanostrip, as shown in Fig. 3(a). To excite all of the coupled modes in this five-skyrmion system, we applied a pulse field only to the left region marked by the rectangular box (*i.e.*, sky1). The FFTs of the temporal oscillations of the $<m_z>$ of the individual skyrmions showed characteristic spectra for the individual skyrmions, as shown in Fig. 3(c). Five peaks denoted as $\omega_i$ ($i$=1, 2, 3, 4, 5) were found, the relative FFT powers of which differed by each skyrmion. For all of the skyrmions, the five different peaks were located at 19.34, 20.31, 21.66, 23.09, and 24.24 GHz. The first and fifth skyrmions showed all five peaks; the second and fourth skyrmions had no third peak, and the third skyrmion had no second or fourth peak. In order to understand the collective breathing modes, we performed, for all of the skyrmions, inverse FFTs of each peak. Figure 3(d) shows the spatial profiles of the $<\Delta m_z>$ variation of the five individual cores for the different coupled modes. The $<\Delta m_z>$ variation profiles represent five different standing-wave modes with fixed boundaries at both ends (the imaginary 0th and 6th skyrmion positions). The shapes of all of the standing-wave modes were either symmetric or anti-symmetric with respect to the center. For the $\omega_1$ mode, all of the skyrmion cores oscillated in anti-phase between the nearest-neighboring skyrmions. For the $\omega_2$, $\omega_3$, and $\omega_4$ modes, the phase difference between the neighboring skyrmions decreased as the wavelength of the standing waves increased, thus resulting in a smaller number of nodes in the standing waves. For the $\omega_5$ mode, all of the cores oscillated in-phase without standing-wave nodes. Accordingly, on the basis of the fixed boundary condition[29], the wave vector of the allowed modes is expressed as $k = (N+1-m)\pi / [(N+1)d_{int}]$, with $N$ the number of skyrmions, $d_{int}$ the skyrmion inter-distance, and $m$ a positive integer number that satisfies $m \leq N$. Therefore, the discrete five-modes' k-values of collective skyrmion core breathing are given by $k_m = (6-m)\pi/6d_{int}$, where $m$ = 1,2,3,4,5, as indicative of each mode. The nodes can thus be found at the $n^{th}$ skyrmion according to the condition $\sin(((6-m)\pi/6d_{int})\hat{\mathbf{k}} \cdot nd_{int}\hat{\mathbf{x}}) = 0$.

As noted earlier, the excited breathing mode from the left skyrmion can propagate through the neighboring skyrmion arrays. Thus, coupled breathing modes in skyrmion lattices can be used as information carriers, owing particularly to the strong coupling between the nearest-neighboring skyrmions. Therefore, we further examined a more general system: a one-dimensional (1D) skyrmion lattice strongly coupled in narrow-width strips as shown in Fig. 4(a). We used 25 skyrmions with $d_{int}$ = 32 nm average inter-distance between nearest-neighboring skyrmions in a given nanostrip. The coupled modes were excited by applying a pulse field only to the first skyrmion at the left end. After the local field was turned off, we monitored the dynamic oscillations of $m_z$ averaged over each skyrmion region under free relaxation. From the FFTs of the temporal oscillations of the $<m_z>$ for each of the 25 skrymion core regions, we obtained a dispersion relation within the corresponding reduced zone, as shown in Fig. 4(b). The overall shape of the dispersion curve was concave down; the frequency was highest at $k$ = 0 and lowest at the Brillouin zone, $k = k_{BZ}$. At $k$=0, all of the skyrmion cores moved coherently, while at $k = k_{BZ,}$ they showed anti-phase motions between the nearest-neighboring skyrmions (*i.e.*, the nodes of the standing waves were between the nearest-neighboring skyrmions), as shown in Fig. 4(c). Such concave-down dispersion is characteristic of the coupled breathing modes of skyrmion motions. This dispersion shape is opposite to the concave-up shape of the coupled gyration modes of skyrmions and the vortex gyration modes. The breathing modes of skyrmions exhibit periodic expansion and contraction of the core in its motion.

Next, we studied how the band structure can be controlled with $d_{int}$ in a 1D skyrmion lattice model. Figure 5(a) shows the variation of the dispersion curve with $d_{int}$. Both band width $\Delta\omega$ and angular frequency $\omega_{k=0}$ at $k = 0$ increased with decreasing $d_{int}$. This behavior can be explained by the variation of the interaction energy between the neighboring cores in dynamic motions for different $d_{int}$ values. In Ref. 37, it was reported that the repulsive force between the skrymion cores increases as $d_{int}$ decreases. In turn, the decrease in $d_{int}$ results in the increase of magnetic interaction energies between the neighboring skyrmion cores. On the other hand, the angular frequency $\omega_{k=BZ}$ at $k = k_{BZ}$ does not significantly vary with $d_{int}$, because the neighboring skyrmion cores expand or contract in anti-phase at $k = k_{BZ}$, resulting in rather less interaction between the neighboring cores.

The external field also can change the dispersion of skyrmion lattices. Here, we performed further simulations by applying perpendicular magnetic fields $H_z = +2, +1, -1$ and $-2$ kOe. Figure 6(a) plots the variation of the dispersion-curve with $H_z$. As $H_z$ increases from negative to positive values, the bandwidth decreases, but $\omega_{k=0}$ and $\omega_{k=BZ}$ increase, as shown in Fig. 6(b). This band structure change is caused by the change in skyrmion core profile along the field direction with field strength. For skyrmions with the downward core, the size of the core increases in negative magnetic fields, whereas it decreases in positive magnetic fields. As a result, the eigenfrequency of the breathing mode of the skyrmion changes linearly with the perpendicular field $H_z$, as reported in Ref. 15. We confirmed by micromagnetic simulations that the eigenfrequency of the breathing mode of a single skyrmion of downward core in a square dot changes linearly with the field strength (see the red line in Fig. 6(b)). Therefore, the linear dependences of $\omega_{BZ}$ on $H_z$ are mainly related to the change in the eigenfrequency of the single skyrmions[15,38,39]. Also, the size of the skyrmion core decreases with increasing $H_z$, resulting in a decrease in the interaction energy between the cores and in $\Delta\omega$ as well (see

Fig.7(c)).

Finally, we calculated a technologically important parameter of the propagation speed of coupled breathing-mode signals through a given nanostrip wherein skyrmions are periodically arranged. Figure 7(a) plots the temporal oscillations of the $<\Delta m_z'>=<m_z(t)>-<m_z(\text{ground state})>$ components of the individual skyrmion cores for the 1D skyrmion array ($N=25$) after applying a pulse field only to the first skyrmion. The excited signal from the left end propagates well through the neighboring skyrmions along the nanostrip. The speed of the collective-breathing-mode signal was estimated, using the 1$^{st}$ wave packet's movement along the skyrmion lattice (denoted by the black line), to be about 340 m/s. This value is about three times faster than that of the gyration signal in the same system[23] and more than five times faster than that of the gyration signal in vortex disk arrays[29]. It has been known that the propagating speed in a two-coupled-vortex system is inversely proportional to the energy transfer time $\tau_{ex} = \dfrac{\pi}{|\Delta\omega|} = \dfrac{\pi\kappa}{\omega_0 |(\eta_x - p_1 p_2 \eta_y)|}$, where $\omega_0$ is the eigenfrequency of the vortex gyration, $\kappa$ the stiffness coefficient, $p$ the vortex polarization, and $\eta$ the interaction strength[36]. Accordingly, the collective breathing modes also can offer an advantage in their propagation speed as an information carrier, since the eigenfrequency of the breathing mode (tens of GHz) is much higher than those of other modes such as the skyrmion gyration mode (~1 GHz) and vortex gyration mode (hundreds of MHz). Further, the speed of the breathing-mode signal was also estimated for different values of $d_{int}$ and $H_z$, as shown in Figs. 7(b) and 7(c), respectively. The resultant propagation speeds generally followed the dependence of $\Delta\omega$ on $d_{int}$ and $H_z$, respectively. The most important point here is that the speed of the breathing-mode signal was found to be reliably controllable by the system, specifically by the skyrmion core inter-distance as well as external parameters such as perpendicular magnetic field strength and direction. For example, for further reduction of skyrmion inter-distance, the speed was increased to ~540m/s for the case of $d_{int}$ = 27 nm, and the speed also was readily increased to ~700m/s by applying a perpendicular field of -2 kOe in the $+z$ direction. Such controllability of the signal speed as well as the dispersion curve of collective breathing modes in 1D skyrmion lattices can be implemented in multifunctional microwave and logic devices operating within a frequency range as wide as a few tens of GHz[40].

**IV. SUMMARY**

In summary, we studied coupled breathing modes in 1D skyrmion lattices in thin film nanostrips. The coupled modes and their characteristic concave-down dispersions were found and explained by a standing-wave model with a fixed boundary condition. The dispersion curve of the collective breathing modes of coupled skyrmions was controllable by skyrmion inter-distance and externally applied perpendicular magnetic fields. Moreover, the coupled breathing modes of skyrmions propagated well through neighboring skyrmions, as fast as 200 ~ 700 m/s, which functionality makes them potential information carriers in future information processing devices.

**ACKNOWLEGEMENTS**

This research was supported by the Basic Science Research Program through the National Research Foundation of Korea (NRF) funded by the Ministry of Science, ICT & Future Planning (NRF-2015R1A2A1A10056286).

**Figure Captions**

FIG. 1. (a) Ground-state single skyrmion in square dot. (b) Temporal variation of average $m_z$ component, $<m_z>$, over entire square dot by excitation of breathing mode using sinc-function field $H_z = H_0 \sin[\omega_H(t-t_0)]/[\omega_H(t-t_0)]$, with $H_0 = 10\,\text{Oe}$, $\omega_H = 2\pi \times 50\,\text{GHz}$, and $t_0 = 1\,\text{ns}$. (c) FFT power spectrum, as obtained from fast Fourier Transform (FFT) of $<m_z>$ oscillations. (d) $<\Delta m_z>$ oscillations obtained from inverse FFTs of peak with frequency band $\Delta f = 22.30 - 22.34$ GHz. On the right are perspective-view snapshot images of the local $\Delta m_z$ of the breathing mode at the indicated two representative times.

FIG. 2. (a) Two skyrmions in rectangular dot. (b) Temporal variation of average $m_z$ components $<m_z>$ over each skyrmion region separated by vertical dotted line ($x = 0 - 32$ nm for left skyrmion and $x = 33 – 64$ nm for right one). (c) FFT power spectra, as obtained from FFTs of $<m_z>$ oscillations for both left and right skyrmion regions. (d) Perspective-view snapshot images of higher- and lower-frequency modes at indicated times for one cycle of each mode

FIG. 3. (a) Five skyrmions in nanostrip of indicted dimensions. The black dotted rectangular box (*i.e.* sky1) indicates the region wherein a pulse field was applied for mode excitation. (b) Temporal evolution of $<m_z>$ oscillations of each skyrmion and (c) their FFT spectra. (d) Spatial profiles of coupled breathing modes for each mode, as represented by $<\Delta m_z>$ of each skyrmion

FIG. 4. (a) 1D skyrmion lattice in 40 nm wide, 800 nm long strip. The colors, as indicated by the color bar, correspond to the out-of-plane components of local magnetizations $m_z = M_z / M_s$. (b) Dispersion curve of 1D skyrmion lattice and (c) Spatial profiles of $\Delta m_z$ components of individual skyrmions for coupled breathing modes at $k=0$ and $k=\pi/a$ where a=32 nm

FIG. 5. (a) Dispersion curves of 1D skyrmion chains and (b) angular frequency at $k = k_0$ and $k = k_{BZ}$ for different inter-distances $d_{int}$= 27, 29, 32, 34, 38 nm.

FIG. 6. (a) Dispersion curves of 1D skyrmion chains for different perpendicular fields $H_z$= -2, -1, 0, 1, 2 kOe. (b) Angular frequencies $\omega_{k=BZ}$, $\omega_{k=0}$, and eigenfrequency $\omega_0$ of breathing mode for isolated skyrmion versus $H_z$.

FIG. 7. (a) Contour plot of $\langle \Delta m_z' \rangle$ of individual cores' oscillations with respect to time and distance in whole chain. The bracket indicates the average value of $m_z$ in each skyrmion region and $\langle \Delta m_z(t)' \rangle = \langle m_z(t) \rangle - \langle m_z(\text{ground state}) \rangle$. (b) Propagation speed and $\Delta\omega$ of coupled breathing mode versus (b) $d_{int}$ and (c) $H_z$

**Figures**

**Fig.1**

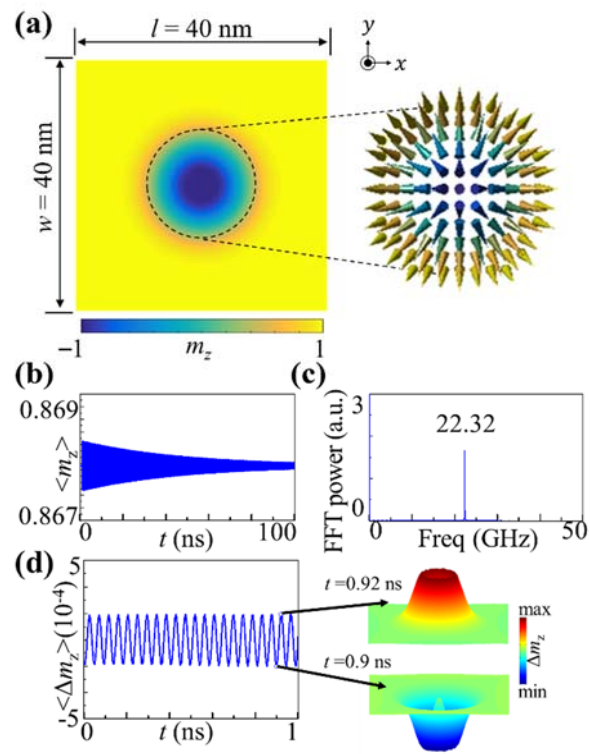

**Fig.2**

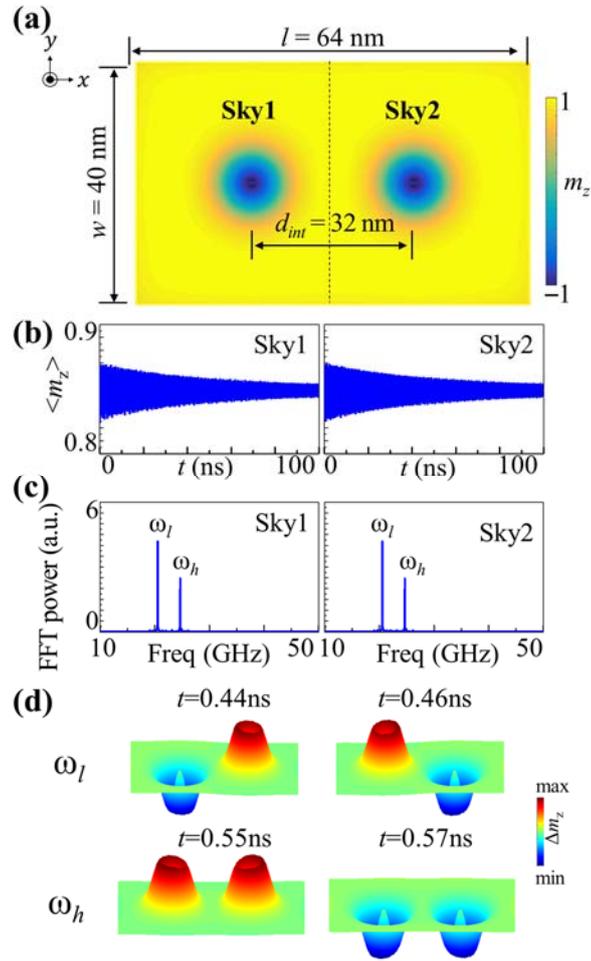

**Fig.3**

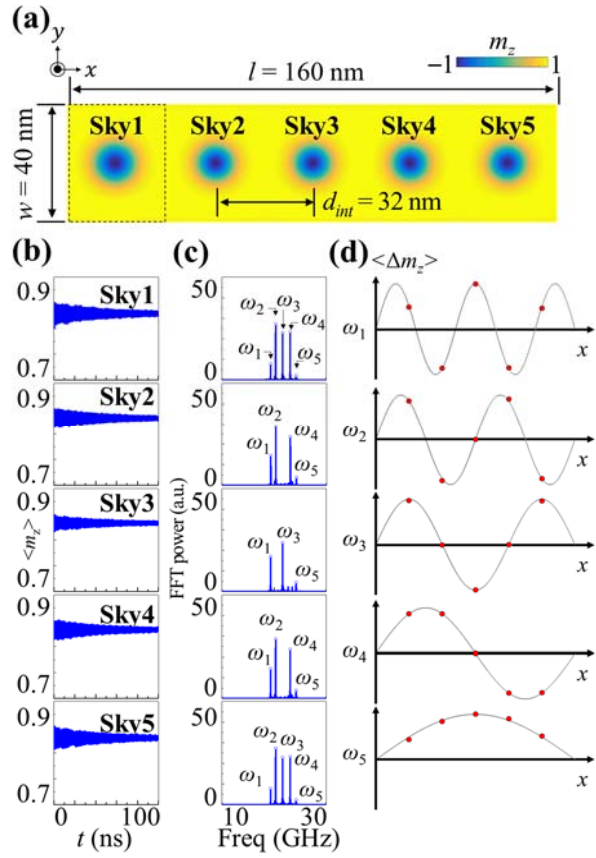

**Fig.4**

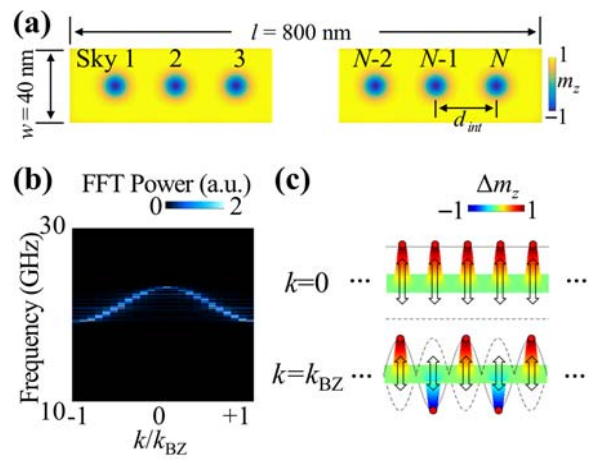

**Fig.5**

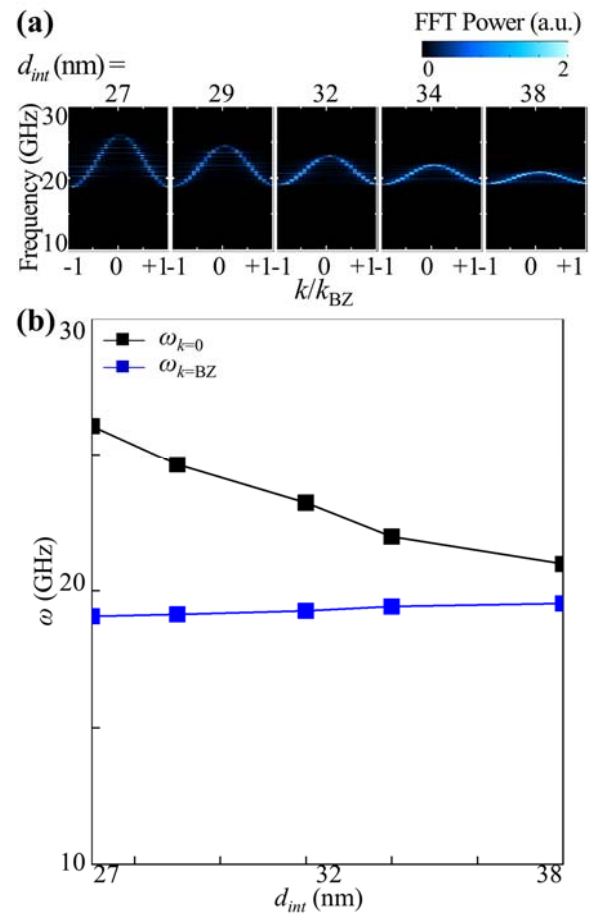

**Fig.6**

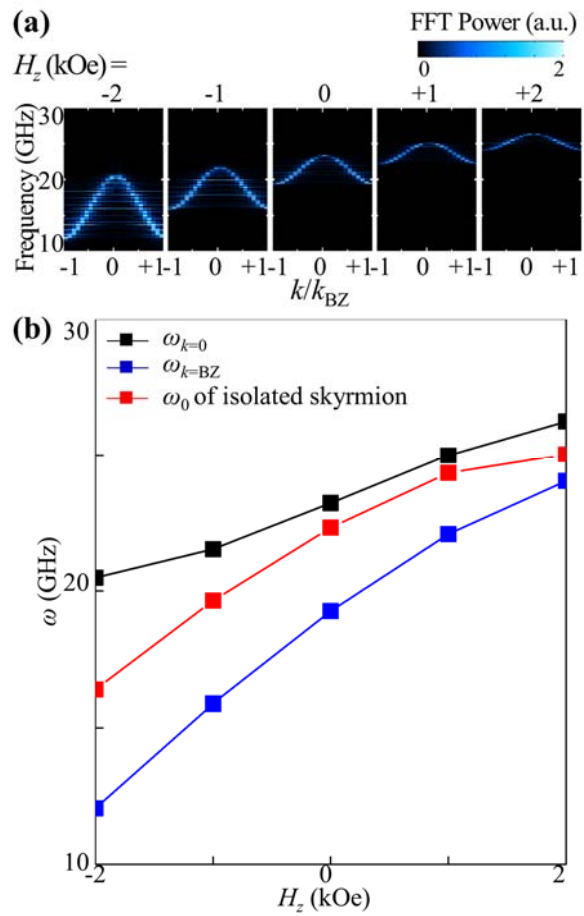

**Fig.7**

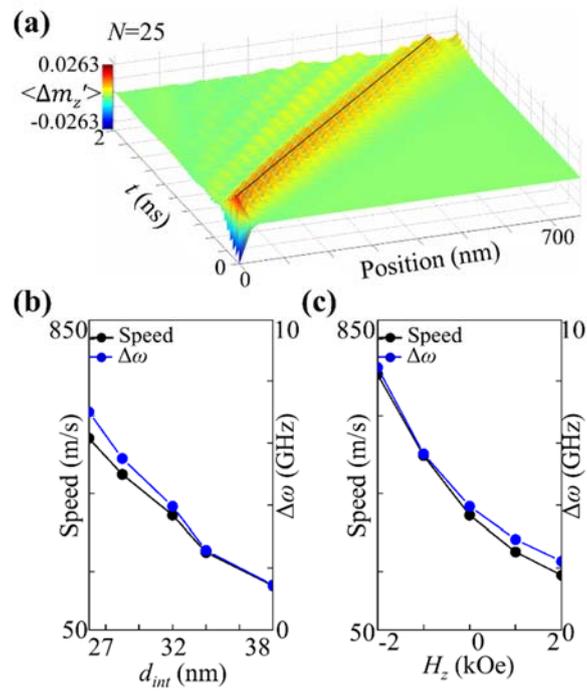